\documentclass{kapproc} 

\usepackage{graphicx}

\normallatexbib

\begin{document}



\articletitle[]{Electronic states and localization in nanoscopic
chains and rings
 from first principles: EDABI method}


\author{}

\author{E.M. G\"orlich$^1$, J.Kurzyk$^2$, A. Rycerz$^1$, R. Zahorbe\'nski$^1$,\\ R.Podsiad\l y$^1$, W. W\'ojcik$^2$,  and J. Spa\l ek$^1$.\\
$^1$ Marian Smoluchowski Institute of Physics, Jagiellonian
University, ulica Reymonta 4, 30-059 Krak\'ow, Poland \\
$^2$ Institute of Physics, Technical University, ulica
Podchora\.zych 1,\\ 30-084 Krak\'ow, Poland }


\begin{abstract}
We summarize briefly the main results obtained within the proposed
{\em EDABI} method combining {\bf E}xact {\bf
D}ia\-go\-na\-li\-zation of (parametrized) many-particle
Hamiltonian with {\em {\bf Ab} {\bf I}nitio} self-adjustment of
the single-particle wave function in the correlated state of
interacting electrons. The properties of nanoscopic chains and
rings are discussed as a function of their interatomic distance
$R$ and compared with those obtained by Bethe ansatz for infinite
Hubbard chain. The concepts of renormalized orbitals, distribution
function in momentum space, and of Hubbard splitting as applied to
nanoscopic systems are emphasized.
\end{abstract}


\section{Introduction}
Recent development in computing techniques, as well as of
analytical methods, has lead to a successful determination of
electronic properties of semiconductors and metals based on LDA
\cite{C1}, LDA+U \cite{C2} and related \cite{C3}  approaches. Even
strongly correlated systems, such as $V_2O_3$ (undergoing the Mott
transition) and high-temperature superconductors, have been
treated in that manner \cite{C4}. However, the discussion of the
metal-insulator transition of the Mott-Hubbard type is not as yet
possible in a systematic manner, particularly for low-dimensional
systems. These difficulties are caused by the circumstance that
the electron-electron interaction is comparable, if not stronger
than the single-particle energy. In effect, the procedure starting
from the single-particle picture (band structure) and including
subsequently the interaction via a {\em local} potential might not
be appropriate then. In this situation, one resorts to
parametrized models of correlated electrons, where the
single-particle and the interaction-induced aspects of the
electronic states are treated on equal footing  (in the Fock space
though)\cite{C5}. The single particle wave functions are contained
in the formal expressions for model parameters. We  propose to
combine
the two efforts in an exact manner, at least for model systems.\\
\indent Our method of approach to the electronic states grew out
of the following question: Can one {\em complete} the procedure
starting from a parametrized model by determining the
single-particle wave functions in the resultant correlated state
{\em a posteriori}? In other words, we determine {\em first} the
energy of interacting particles in terms of the microscopic
parameters rigorously and only then optimize this energy with
respect to the wave functions contained in those parameters by
deriving the {\em self-adjusted wave equation} for this state.
Physically, the last step amounts to allowing the single-particle
wave functions to relax in the correlated state. This method has
been overviewed in number of papers \cite{C6,C7,C8,C9,C10}, so we
present here examples of its application  to low-dimensional and
nanoscopic systems. We start with the analysis of the infinite
Hubbard chain and then compare the results with those for finite
chains. We also discuss briefly small hydrogenic ring of $N=6$
atoms. Also, throughout the paper we are using adjustable Wannier
composed of atomic or Gaussian functions, which are determined
explicitly from the minimization of the system ground state energy
as a function of interatomic distance. The paper describes various
ground-state characteristics of simple monoatomic chains and
rings.
\section{Exact diagonalization combined with wave-function
determination: formal aspects} 
\begin{figure}
\centerline{\includegraphics[width=11cm,height=7.0cm]{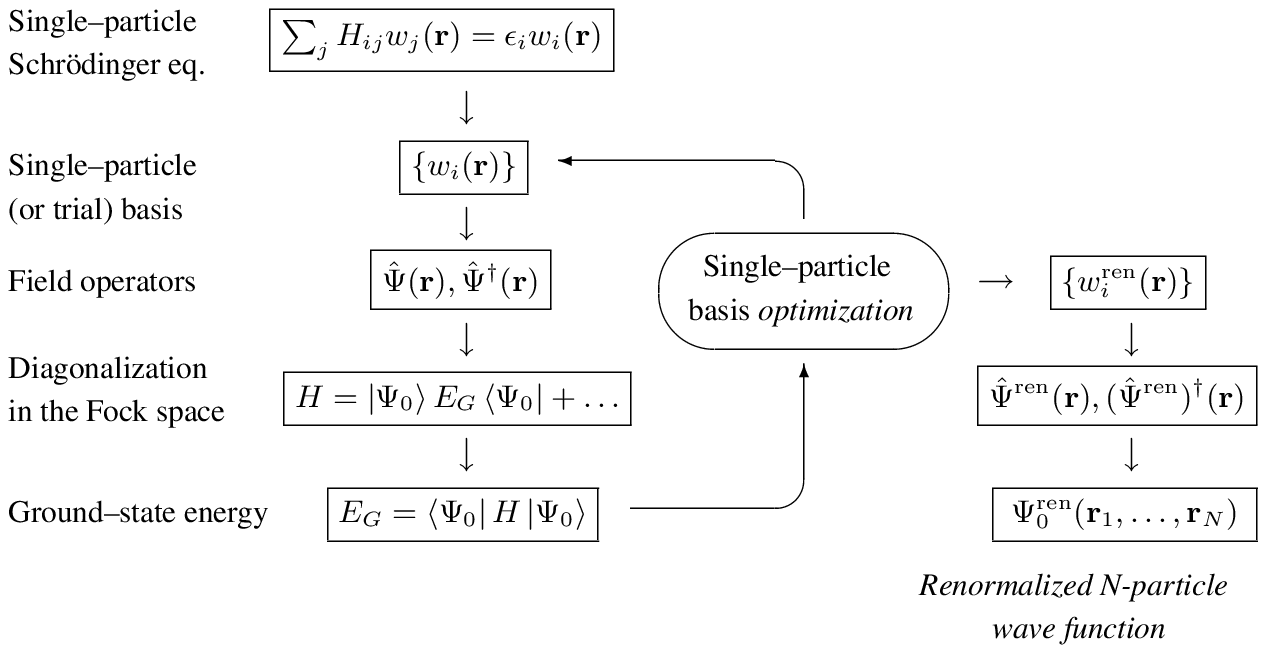}}
\caption{\protect\inx{Flowchart} of the EDABI method described in
this paper. The top line is missing when the adjustable Gaussian
basis is used.} \label{F0}
\end{figure}

As our method contains both
many-particle and single-particle (wave-function) aspects, both
treated in a rigorous manner, it may be useful to summarize the
{\em basic principle} behind it. First, we start with the standard
expression of the many-particle Hamiltonian in the Fock space
\[
\hat H = \int d^3{\bf r} \hat \Psi^\dagger ({\bf r}) H_1({\bf r})
\hat \Psi ({\bf r }) + \]
\begin{equation}
 \frac{1}{2} \int d^3r d^3r' \hat
\Psi^\dagger ({\bf r}) \hat \Psi^\dagger ({\bf r'}) H_2({\bf r -
r'}) \hat \Psi ({\bf r'})\hat \Psi ({\bf r}),
 \label{Ea1}
\end{equation}
where $H_1$ and $H_2$ are the Hamiltonian for one and one pair of
particles, and
\begin{equation}
\hat \Psi ({\bf r}) = \sum_{i} w_i ({\bf r}) {a_{i \uparrow}
\choose a_{i \downarrow}} \equiv \sum_{i} w_i ({\bf r}) a_i,
\label{Ea2}
\end{equation}
is the field operator, $\{ w_i({\bf r}) \}$ is the single-particle
basis of wave-functions (complete, but otherwise {\em arbitrary}),
ad $a_{i \sigma}$ is the annihilation operator of the particle in
the single-particle state $|i \sigma >$ represented by $w_i({\bf
r})$ and the spin quantum number $\sigma = \pm 1$. The only
approximation we make in our whole analysis is that instead of
taking the summation over a complete set $\{i\}$ of
single-particle states (and transition between them), we limit
ourselves to a finite subset of $M$ states. This means that we are
solving a {\em model many-body system} rather than the complete
problem ({\em Hubbard} or {\em extended Hubbard models} are
classic
examples representing one-orbital-per-atom).\\
\indent The essential step in our analysis follows from taking the
finite single-particle basis, which amounts to limiting the
occupation-number representation space to a space of finite
dimension. To minimize the error in estimating the ground-state
energy of many-particle system we calculate first {\em all}
configurations in the limited Fock subspace and then optimize the
orbitals in the interacting (correlated) ground state. In this
manner, the second quantization takes care of counting various
many single-particle microconfigurations enforced by the
interaction between them, whereas the wave-function optimization
adjusts each of them to the milieu of all others. In brief,
second-quantization aspect addresses the question {\em how} they
are distributed among the single-particles state and the
first-quantization optimization
tells us how their states look like once they are there.\\
\indent  One should also address the problem of many-body vs.
single-particle wave function. In the wave mechanics of the
interacting system only the $N$-particle wave function $\Psi ({\bf
r}_1, \ldots {\bf r}_N)$ has a sense. However, in the second
quantization the  single-particle wave function appears explicitly
in the expression for the field operator, the remaining part is
the evaluation of various microconfigurations, with proper weights
characterized by their energy (no entropy appears as they form a
single coherent state). In other words, the microconfiguration
counting in the occupation-number representation replaces the
determination of $N$-particle Hilbert space. But then, we are
faced with the single-particle states determination, on which the
counting is performed. Explicitly, the $N$-particle state $|
\Phi_0
>$ in the Fock space can be defined as
\begin{equation}
\vert \Phi_0 \rangle = \frac{1}{\sqrt{N!}} \int d^3{\bf r}_1 \dots
{\bf r}_N \Psi_0 ({\bf r_1, \dots , r_N}) \hat \Psi^\dagger ({\bf
r_1}) \dots \hat \Psi^\dagger ({\bf r_N}) \vert 0 \rangle,
\label{Ea3}
\end{equation}
where $\vert 0 \rangle$ is the vacuum state. The $N$-particle wave
function is then determined from
\begin{equation}
\Psi_0 ({\bf r_1, \dots , r_N}) = \frac{1}{\sqrt{N!}} \langle 0
\vert \hat \Psi ({\bf r_1}) \dots \hat \Psi ({\bf r_N}) \vert
\Phi_0 \rangle.
\label{Ea4}
\end{equation}
Expanding $| \Phi_0 >$ in the basis involving $M$ single-particle
states, i.e.
\begin{equation}
\vert \Phi_0 \rangle = \frac{1}{\sqrt{N!}} \sum_{j_1, \dots, j_N =
1}^{M} C_{j_1 \dots j_N} a^\dagger_{j_1} \dots a^\dagger_{j_N}
\vert 0 \rangle, \label{Ea5}
\end{equation}
we obtain the $N$-particle wave function in the form
\[
\Psi_0 ({\bf r_1, \dots r_n}) = \frac{1}{N!} \sum_{i_1, \dots,i_N
= 1}^M  \sum_{j_1, \dots, j_N = 1}^{M} \langle 0 \vert a_{i_N}
\dots a_{i_1} a^\dagger_{j_1} \dots a^\dagger_{j_N} \vert 0
\rangle \] \begin{equation}C_{j_1 \dots j_N} w_{i_1}({\bf r_1})
\dots w_{i_N}({\bf r_N}). \label{Ea6}
\end{equation}
\indent  The many-body coefficients $C_{j_1 \dots j_N}$ will be
calculated from either the direct diagonalization or the Lanczos
algorithm, whereas the wave functions $\{ w_i({\bf r}) \}$ will be
determined from the {\em renormalized (self-adjusted) wave
equation}, which is set up in the following manner. First, we
substitute (\ref{Ea2}) into (\ref{Ea1}) and obtain the formal
expression for the ground state energy
\begin{equation}
E_G \equiv \langle H \rangle = \sum_{i j \sigma} t_{i j} \langle
a^\dagger_{i \sigma} a_{j \sigma} \rangle + \frac{1}{2} \sum_{i j
k l \sigma_1 \sigma_2} V_{i j k l} \langle a^\dagger_{i \sigma_1}
a^\dagger_{j \sigma_2} a_{l \sigma_2} a_{k \sigma_1} \rangle,
\label{Ea7}
\end{equation}
where the microscopic parameters
\[
 t_{i j} = \int d^3{\bf r}
w^\star_i({\bf r}) H_1({\bf r}) w_j({\bf r}),
\]
 and
\[
V_{i j k l} = \int d^3{\bf r_1} d^2d^3{\bf r_2} w^\star_i({\bf
r_1}) w^\star_j({\bf r_2}) V({\bf r_1 - r_2}) w_k({\bf r_1})
w_l({\bf r_2}),
\]
contain the single-particle wave functions and the averages are
$\langle a^\dagger_{i \sigma} a_{j \sigma} \rangle \equiv \langle
\Phi_0 | a^\dagger_{i \sigma} a_{j \sigma} | \Phi_0 \rangle$, etc.
Second, in the situation when we work with definite number of
particles  (or else, if the chemical potential can be regarded as
constant), then we can determine $\{ w_i({\bf r}) \}$ by setting
Euler equations for each of them, i.e. by minimizing the
functional $F=E_G\{w_i({\bf r}),\nabla w_i({\bf r}) - \sum_i
\lambda_i \int d^3 {\bf r} w_i^\star({\bf r} w_i({\bf r}$. Such a
procedure leads to the equation
\begin{equation}
\frac{\delta E_G}{\delta w^\star_i({\bf r})} - \nabla \frac{\delta
E_G}{\delta (\nabla w^\star_i({\bf r}))} = \lambda_i w_i({\bf r})
\label{Ea8}
\end{equation}
A direct solution of this equation is very difficult to achieve.
Therefore, we define the starting wave functions $w_i^{(0)}({\bf
r}) = \sum_{j=1}^M \beta_{ij} \Phi_j({\bf r}; \alpha)$, where
$\beta_{ij}$ are the mixing coefficients and $\Phi_j({\bf r};
\alpha)$ are the atomic wave functions of the size $\alpha^{-1}$.
In result,  the renormalized wave equation reduces to the
minimization of (\ref{Ea7}) with respect to $\alpha$ (there is
only one size $\alpha^{-1}$ when we take orbitals of the same type
for each atomic site in the system). The whole EDABI procedure is
schematically summarized in Fig. \ref{F0}.
\section{Electron states for the Hubbard chain and a comparison with nanochains} 
\begin{figure}
\centerline{\includegraphics[width=10cm,height=8cm]{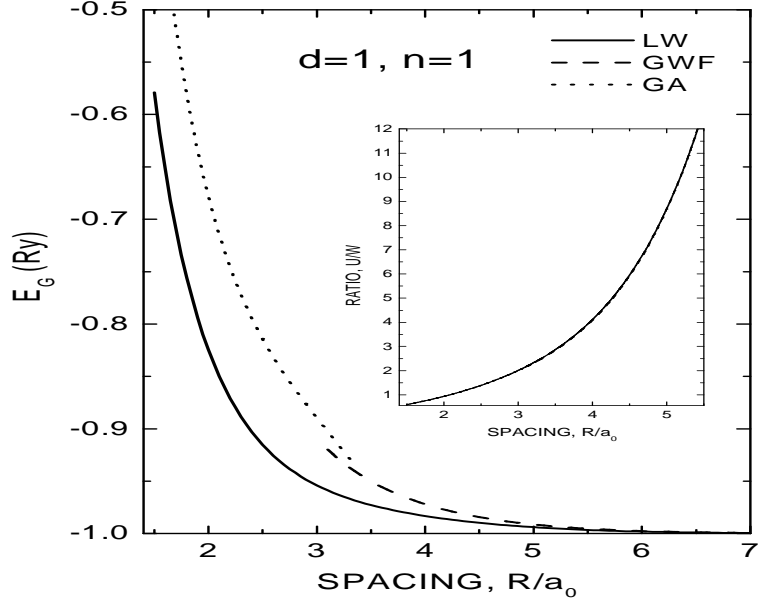}}
 \caption{\protect\inx{Optimized ground state energy} as a function
of interatomic distance - solid line, Gutzwiller ansatz(GA)
solution - dotted line, Gutzwiller wave-function (GWF) - dashed
line. }
 \label{F1}
 \end{figure}

\begin{figure}
\centerline{\includegraphics[width=10cm,height=8cm]{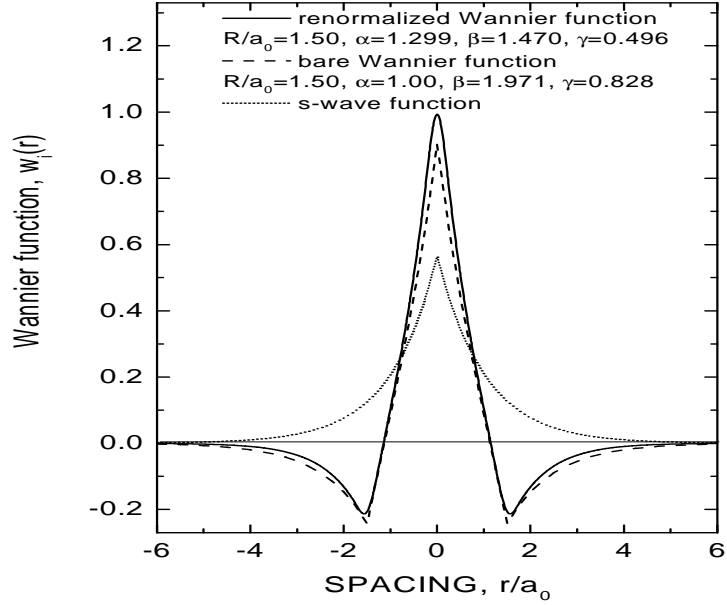}}
\caption{\protect\inx{The shape of} the optimized Wannier function
- solid line, non-optimized Wannier function - dashed line, 1s
atomic wave function - dotted line (for comparison).}
\label{F2}\end{figure}

We implement
first the EDABI method to the case of the linear chain composed of
$N$ atoms, which obeys periodic boundary conditions. In the
simplest situation we have one valence electron per atom, as is in
the case of monoatomic chains composed of Na, K or Cs. However,
for the sake of simplicity, we consider here the chain composed of
hydrogen atoms; this means only that we compose the Wannier
function of the conduction band from 1s-like atomic states of an
adjustable size (there is no principal problem in considering
$n$s-like states,
with $n>1$).\\
\indent As mentioned earlier, we start from the many-body model of
interacting electrons. For this purpose, we consider an extended
Hubbard model, as represented by the Hamiltonian
{
\[ H =
\epsilon_a \sum_{i\sigma} n_{i \sigma}+t \sum_{i\sigma} \left(
a^\dagger_{i \sigma} a_{i+1 \sigma} + h.c. \right)
\]
\begin{equation}
 + U \sum_i
n_{i \uparrow} n_{i \downarrow}+ \sum_{i<j} K_{ij} n_i n_j +
\sum_{i<j} V_{ion}(R_i-R_j) \label{E1}
\end{equation}}
The first term represents the atomic energy ($\epsilon_a \equiv
\langle w_i | H_1 | w_i \rangle$, where $H_1$ is the Hamiltonian
for a single particle in the system and $w_i\equiv w_i({\bf r})$
is the Wannier state centered on site $i$ in that system). The
second term is the so-called hopping term with the hopping
integral $t \equiv \langle w_i | H_1 | w_{i\pm 1} \rangle$(we use
the tight-binding approximation and disregard more distant
hoppings). The  next two terms represent respectively  the intra-
and inter- atomic parts of the Coulomb interaction between
electrons, with $K_{ij} \equiv \langle w_i w_j | V_{12} | w_i w_j
\rangle$, $U=K_{ii}$, and where $V_{12}({\bf r} - {\bf r}')$ is
the classical Coulomb interaction for pair of electrons. The last
term expresses the classical repulsion between the ions located at
sites $i$ and $j$. Also, the single particle operator $H_1$
contains both kinetic energy and the Coulomb attractive
interaction of ions (it is sufficient to take $6\div 10$ ionic
Coulomb wells surrounding the electron located on site i to
reproduce to a good accuracy the effective values od $\epsilon_a$ and $t$).\\
 \indent For a detailed analysis it is
convenient to rewrite (\ref{E1}) in the equivalent form, which for
the case of one electron per atom reads
\begin{equation}
H = \epsilon_a^{eff} N_e + t \sum_{i\sigma}\left( a^\dagger_{i
\sigma} a_{i+1 \sigma} + h.c. \right) + U \sum_i n_{i \uparrow}
n_{i \downarrow} + \sum_{i<j} K_{ij} \delta n_i \delta n_j ,
\label{E2}
\end{equation}
where $\delta n_i \equiv 1 - n_i$, $N_e = \sum_{i\sigma}
n_{i\sigma}$ is the number of electrons (here equal to the number
of atoms), and
\begin{equation}
\epsilon_a^{eff}\equiv\epsilon_a + \frac{1}{N}\sum_{i<j}\left(
K_{ij}+\frac{e^2}{|R_i-R_j|} \right)  \label{E3}
\end{equation}
is the effective atomic energy, which includes the compensating
repulsive interactions to provide correctly the atomic limit. We
disregard the last term in (\ref{E2}), since we would like to
relate the nanochain results to those for the Hubbard chain for $N
\rightarrow \infty$ \cite{C11}. Under these circumstances,
Hamiltonian (\ref{E2}), we consider here explicitly, represents
only the Hubbard model with the energy part $\epsilon_a^{eff}$
providing a proper neutral-atom limit for $R \rightarrow \infty$.
Eq. (\ref{E2}) then can be diagonalized exactly with the help of
Bethe ansatz \cite{C12}. Explicitly, the expression for the ground
state energy in terms of microscopic parameters
$\epsilon_a^{eff}$, $t$, and $U$ takes the form
\begin{equation}
\frac{E_G}{N} = \epsilon_a^{eff}+4 t \int_0^\infty d\omega
\frac{J_0(\omega) J_1(\omega)}{\omega [ 1 + \exp ( -\omega U / 2
t)]}, \label{E4}
\end{equation}
where $J_n(x)$ is the Bessel function. One should note that the
energy expression is not an additive function of terms $\sim t$
and $\sim U$, as the solution is of nonperturbative nature.\\
\indent As we have stressed earlier, the Lieb-Wu solution
(\ref{E4}) does not represent the final step of the analysis as it
still contains parameters, which in turn are expressed through the
one-particle (Wannier) functions $\{ w_i({\bf r}) \}$. Therefore,
we minimize the energy {\em functional} $E\equiv E\{  w_i({\bf r})
, \nabla w_i({\bf r}) \}$ with respect to $\{  w_i({\bf r}) \}$.
Such a procedure leads to the Euler variational principle for {\em
renormalized} or {\em self-adjusted} wave functions in the
correlated state \cite{C5}. Only after solving that equation and
calculating explicitly the parameter values for given $R$ we
obtain the energy of the correlated ground state as a function of
the lattice parameter.
This last step completes the theoretical analysis of the Hubbard model.\\
\indent Technically, we compose the Wannier functions of atomic
1s-like functions of variable size, with respect to which we
minimize $E$. Explicitly, in the spirit of tight-binding
approximation we can write that
\begin{equation}
w_i({\bf r}) = \beta \psi_i({\bf r}) - \gamma [ \psi_{i+1}({\bf
r}) + \psi_{i-1}({\bf r}) ], \label{E5}
\end{equation}
where $\beta$ and $\gamma$ are the mixing coefficients determined
from the orthonormality condition $\langle w_i | w_j \rangle =
\delta_{ij}$ and
\begin{equation}
\psi_i({\bf r}) = (\alpha^3/\pi)^{1/2} \exp( - \alpha |{\bf r} -
{\bf R}_i| ), \label{E6}
\end{equation}
with $\alpha$ being the adjustable parameter. Substituting
(\ref{E5}) to the expressions for  $\epsilon_a$, $t$, and $U$, and
then those expressions to (\ref{E4}) we obtain the physical ground
state energy as the minimal energy with respect to $\alpha$ for
given lattice constant $R\equiv |R_i-R_{i\pm 1}|$. \\
\indent In Fig. \ref{F1} we plot the optimized ground state energy
with respect to $\alpha$ as a function of interatomic distance. We
have also added there the corresponding values obtained using the
Gutzwiller ansatz (GA)\cite{C13} and the Gutzwiller wave-function
(GWF)\cite{C14} approximations. Obviously, the approximate
solutions \cite{C13,C14} represent the upper estimates for the
system energy. In the inset we have plotted the
interaction-to-bandwidth ratio $U/W$; one sees that the electrons
are strongly correlated (i.e. $U/W > 1$) for realistic values of
$R$ ($a_0 \simeq 0.53$ \AA \ is the Bohr radius). Obviously, the
chain composed of hydrogen atoms is not stable, as the energy
$E_G$ (per atom) is above $-1$Ry. However, this is not an issue
here, since we are considering only a model situation (such a
chain could be stabilized on  a substrate, but then  we should
have to include the trapping potential, in addition to the
periodic potential composed of the protonic Coulomb wells). It
would be interesting to repeat this analysis for 2s and 3s
orbitals representing the conduction band of a quantum wire
composed of Li and Na respectively; this
does not present a major obstacle.\\
\indent In Fig. \ref{F2} we draw the Wannier function centered on
site "0" for the hydrogen chain: the solid line represents the
self-adjusted Wannier function in the tight-binding approximation,
whereas the dashed curve is the usual Wannier function calculated
in TBA. For comparison, the dotted line is the usual 1s-type
atomic function. All the curves were drawn along the chain
direction. Although the differences in the first two cases do not
seem crucial, the values of microscopic parameters:
$\epsilon_a^{eff}$, $t$, $U$ and $K$ differ remarkably. The
determined values of those parameters versus $R/a_0$ are provided
in Table \ref{T1} (the values of $E_G$ and $\alpha$ are also
listed there). One notices numerically, that
$U/W>1$, where $W=4|t|$, for $R \geq 2 a_0$.\\
\begin{table}
\begin{center}
\begin{tabular}{rclllll}
\hline\hline \it $R/a_0$ & $\alpha_{\min}a_0$
 & $\epsilon^{eff}_a$ & $t$ & $U$ & $K$ & $E_G/N$ \\
\hline 1.5 & 1.806 &
  0.9103 &  -1.0405 & 2.399 & 1.695 &  0.0665 \\
2.0 & 1.491 &
  -0.1901 & -0.5339 & 1.985 & 1.172 & -0.5179 \\
2.5 & 1.303 &
  -0.6242 & -0.3076 & 1.722 & 0.889 & -0.7627 \\
3.0 & 1.189 &
  -0.8180 & -0.1904 & 1.553 & 0.713 & -0.8800 \\
3.5 & 1.116 &
  -0.9104 & -0.1230 & 1.440 & 0.596 & -0.9391 \\
4.0 & 1.069 &
  -0.9559 & -0.0815 & 1.365 & 0.513 & -0.9693 \\
4.5 & 1.039 &
  -0.9784 & -0.0546 & 1.317 & 0.451 & -0.9848 \\
5.0 & 1.022 &
  -0.9896 & -0.0370 & 1.288 & 0.403 & -0.9926 \\
6.0 & 1.013 &
  -0.9977 & -0.0165 & 1.269 & 0.334 & -0.9982 \\
7.0 & 1.001 &
  -0.9995 & -0.0072 & 1.252 & 0.286 & -0.9996 \\
8.0 & 1.001 &
  -0.9999 & -0.0031 & 1.251 & 0.250 & -0.9999 \\
10.0 & 1.000 &
  -1.0000 &  0.0003 & 1.250 & 0.200 & -1.0000 \\
\hline\hline
\end{tabular}
\end{center}
\caption{Optimized inverse orbital size, microscopic parameters
and the ground--state energy for $N=10$ atoms calculated in
Slater--type basis, as a function of interatomic distance.
Intersite Coulomb repulsion $K_1$ is included on the mean--field
level in $\epsilon^{eff}_a$, Hubbard $U$ term is treated exactly.
Single--particle potential contains \emph{six} Coulomb wells.}
\label{T1}
\end{table}
\indent The detailed electronic properties of the chain have been
discussed separately \cite{C15}. Probably, the most interesting
result of relevance to this workshop is the conclusion that the
values of the energy per atom (and of microscopic parameters as
well) are almost the same when obtained either from the solution
for $N \rightarrow \infty$ (discussed above) and from the
numerical solution for, say, $N=10$ atoms \cite{C16}. This
statement is illustrated in Table \ref{T2}, where the optimal
inverse size $\alpha_{min}$ of the atomic wave functions composing
$w_i({\bf r})$ and that of ground-state energy have been listed as
a function the interatomic distance. The three columns in each
category represent respectively the following results: (i) when
single-site and two-site parameters have been calculated for
1s-like functions and 3- and 4-site terms (in the atomic basis)
have been calculated in the contracted STO-3G basis; (ii) and
(iii) represent {\em all} calculations in the Gaussian basis. The
results are pretty close, independently of the trial atomic basis
selected to represent $w_i({\bf r})$. However, there is one
restriction, namely the boundary conditions selected in the $N
\rightarrow \infty$ and $N =10$ cases must coincide ( here they
are selected as periodic b.c.). The importance of the results
displayed in Table \ref{T2} relies on the fact that in this manner
we can apply analytic results obtained for the infinite chain to
the finite chains (nanowires) if only realistic single particle
wave functions are taken into account. Obviously, in Table
\ref{T2} the comparison was made for the case of one electron per
atom only (the Lieb-Wu solution provides then the insulating
state). We plan calculating the properties of Li and Na nanowires
starting from this prescription. Parenthetically, since in the
infinite chains we have {\em charge-spin separation} and other
{\em non-Fermi liquid effects} \cite{C17}; they should appear also
in some form in non-half filled nanoscopic chains, which should
also be regarded as strongly correlated systems from the start.
This last question will be taken up again in Sec.5.\begin{table}
\begin{center}
\begin{tabular}{ c c c c || c c c }
\hline \hline & & $\alpha_{min}$ &
& & $E_G$&  \\
\cline{2-7}
 $R/a_0$ & $N=\infty$  &  $N=\infty$  & $N=10$ & $N=\infty$  &
$N=\infty$  &  $N=10$ \\
& 1s/3G  & 3G & 3G & 1s/3G  & 3G  & 3G \\
\hline
 1.5 &1.2985 &1.3062 &1.3094 &-0.5788 &-0.5527 &-0.5684\\
 2.0 &1.1753 &1.2038 &1.2047 &-0.8246 &-0.8104 &-0.8154\\
 2.5 &1.0924 &1.1203 &1.1203 &-0.9152 &-0.9123 &-0.9139\\
 3.0 &1.0485 &1.0683 &1.0672 &-0.9540 &-0.9560 &-0.9567\\
 4.0 &1.0212 &1.0212 &1.0203 &-0.9832 &-0.9840 &-0.9841\\
 5.0 &1.0109 &1.0062 &1.0054 &-0.9939 &-0.9900 &-0.9901\\
 6.0 &1.0055 &1.0020 &1.0027 &-0.9981 &-0.9914 &-0.9914\\
 7.0 &1.0021 &1.0005 &1.0000 &-0.9995 &-0.9917 &-0.9917\\
 8.0 &1.0005 &1.0001 &1.0000 &-0.9999 &-0.9917 &-0.9917\\
 10.0 &1.0002 &1.0001 &1.0000 &-1.0000 &-0.9917 &-0.9917\\
\hline \hline
\end{tabular}
\end{center}
\caption{Renormalized values of the inverse size ($\alpha$) of the
atomic wave-function (columns $2 \div 4$) and the corresponding
values of the ground state energy (columns $5 \div 7$), both
versus the lattice parameter $R$.} \label{T2}
\end{table}
\section{Nanoscopic $H_N$ rings}
\begin{figure}
\centerline{\includegraphics[width=10cm,height=8cm]{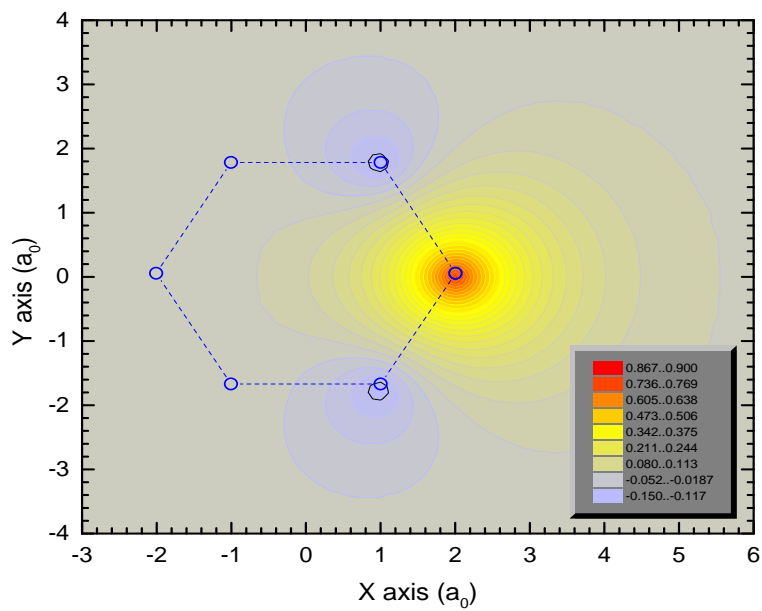}}
\caption{\protect\inx{Spatial} wave-function profiles for a
selected site of $H_6$ cluster of hexagonal shape. Note the
negative value on the neighboring site to the central atom.}
\label{F3}
\end{figure}

\begin{figure}
\centerline{\includegraphics[width=10cm,height=8cm]{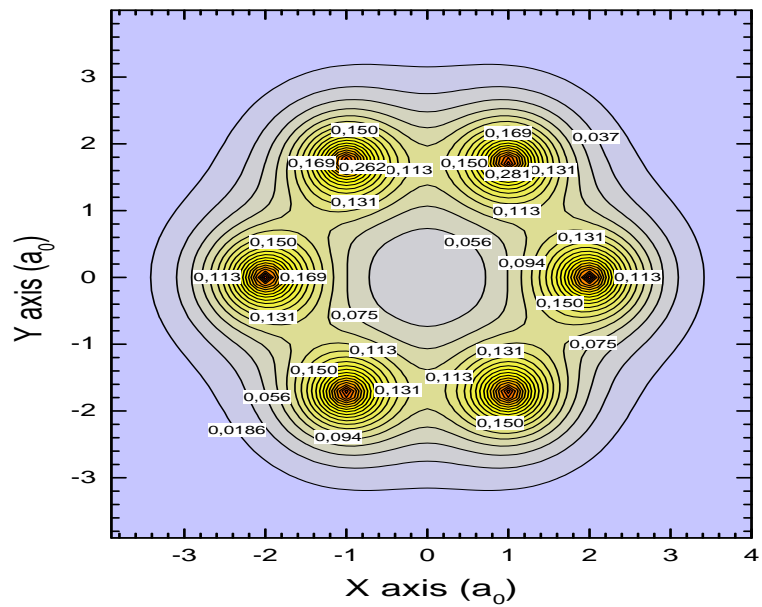}}
\caption{\protect\inx{Exact} density profiles for electrons in a
hexagonal ring of atoms. } \label{F4}
\end{figure}

\begin{figure}
\centerline{\includegraphics[width=10cm,height=7cm]{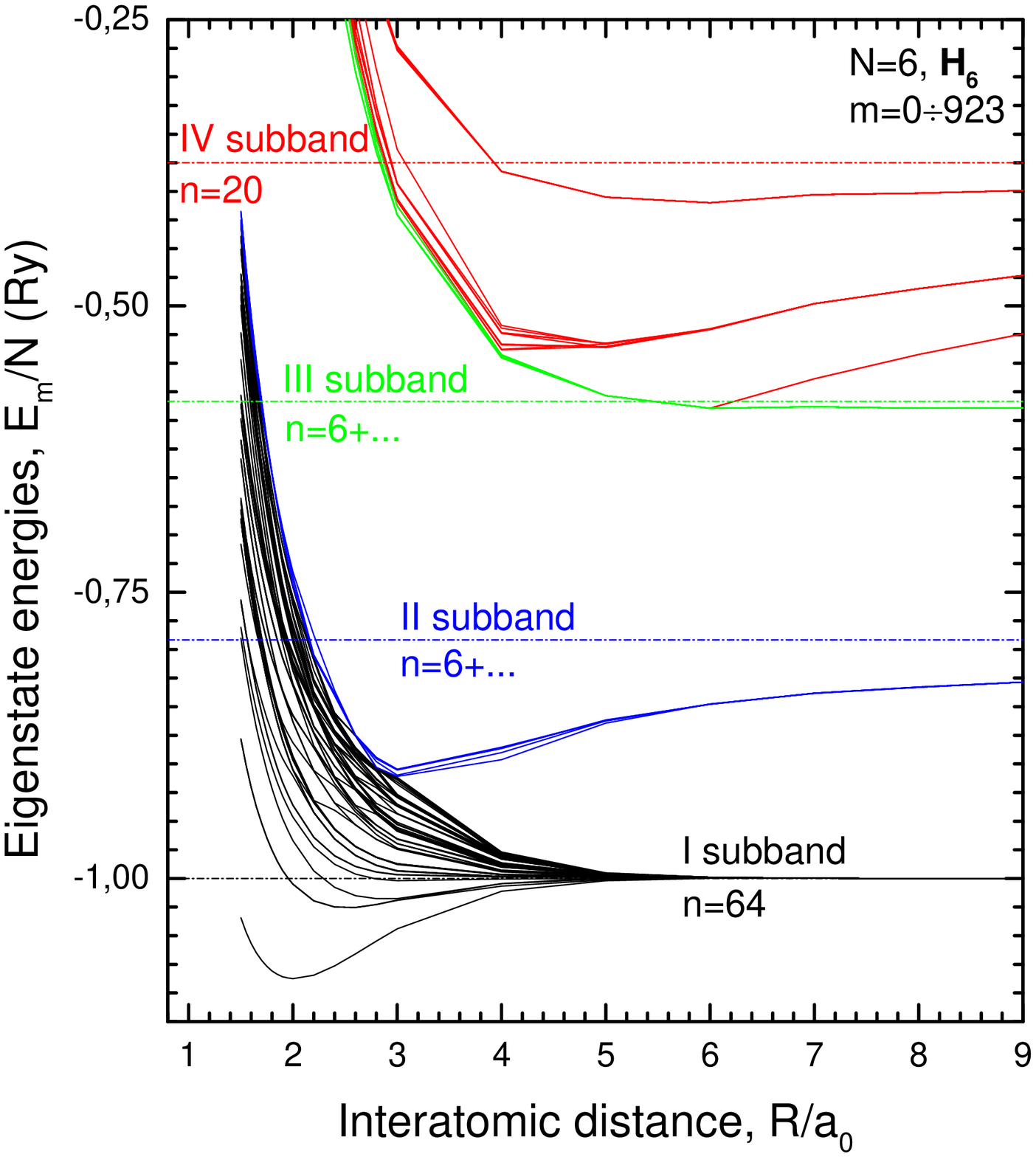}}
\caption{\protect\inx{Decomposition} of the system energies into
Hubbard subbands for $H_6$ cluster, plotted as a function of
interatomic distance. The horizontal lines represent the atomic
limit values for the levels $\frac{U l}{N}$, with $l = 0,1,2,$ and
$3$.} \label{F5}
\end{figure}

As a second example we consider a ring composed of $N=6$ hydrogen
atoms arranged planarly (the stable $H_4$ clusters arranged
spatially have been considered elsewhere \cite{C16,C5}). In this
situation, the periodic boundary conditions (PBC) are the {\em
physical} condition for the system geometry. In Fig. \ref{F3} we
plot the profile of the wave function located around the exemplary
atom in the hexagon $H_6$. This density contains renormalized
orbitals and the calculations of the ground states involve (in
principle) ${12 \choose 6} = 924$ 6-particle states in the
occupation number representation spanned on $M=12$ states and
containing 6 Wannier functions of adjustable size. The space
profiles are useful for the determination of the density function
profile $n({\bf r}) \equiv \langle \hat{\Psi}^\dagger({\bf r})
\hat{\Psi}({\bf r}) \rangle$, where $\hat{\Psi}({\bf r})$ is the
field operator spanned on those 6 Wannier states. In effect, we
have that
\begin{equation}
n({\bf r}) = \sum_{i\sigma} |w_i({\bf r})|^2 \langle n_{i\sigma}
\rangle + {\sum_{ij\sigma}} ' w_i^\star({\bf r})w_j({\bf r})
\langle a^\dagger_{i\sigma}a_{j\sigma} \rangle, \label{E7}
\end{equation}
where the primed summation means that $i \neq j$. The second part
represents the part which does not appear in the Hartree-Fock
(single determinant) approximation for the many particle wave
function. In Fig. \ref{F4} we display the electron density
profiles (normalized to unity) for the $N = 6$ atoms arranged in
the hexagon,
 as is also  in the translationally invariant along the ring density profile $n({\bf r})$.\\

\indent An interesting feature of the spectrum of electronic
states arises when the distance between the atoms increases.
Namely, the spectrum decomposes into well defined Hubbard
subbands, as shown in Fig. \ref{F5} (more appropriately, they
represent {\em manifolds} corresponding to the {\em subbands} when
$N \rightarrow \infty$). The lowest manifold (I) corresponds to
the configuration with approximately singly occupied orbitals
(highest occupied Wannier orbitals) whereas the manifolds II-IV
correspond respectively to the states with one to three double
occupancies. This division into the well separated manifolds for
larger $R$ is even better seen for the clusters of $N=4$ and 5
atoms \cite{C8}. One should mention that the states considered
here represent the excited states calculated with the help of
Lanczos procedure \cite{C5}, repeated many times until the
configuration with the minimal energy and the optimal
single-particle wave function size are reached simultaneously.
\section{Further features of the results: collective properties of nanochains}
\subsection{One electron per atom case: localization threshold}
\begin{figure}
\centerline{\includegraphics[width=10cm,height=7cm]{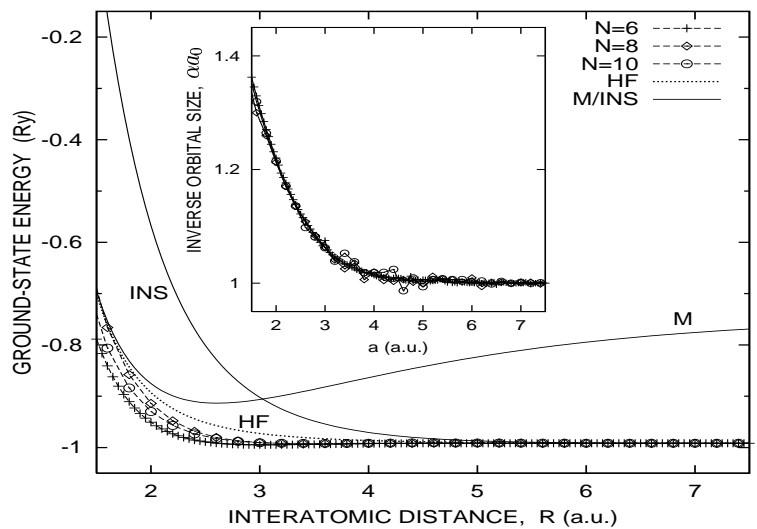}}
\caption{Ground state energy per atom vs. R for the linear chain
with $N = 6 \div 10$ atoms with periodic boundary conditions. The
STO-3G Gaussian basis for representation of atomic orbitals
forming the Wannier function has been used. The inset provides a
universal behavior of the inverse size $\alpha$ of the orbitals.
For details see main text.} \label{F5a}
\end{figure}

\begin{figure}
\centerline{\includegraphics[width=10cm,height=7cm]{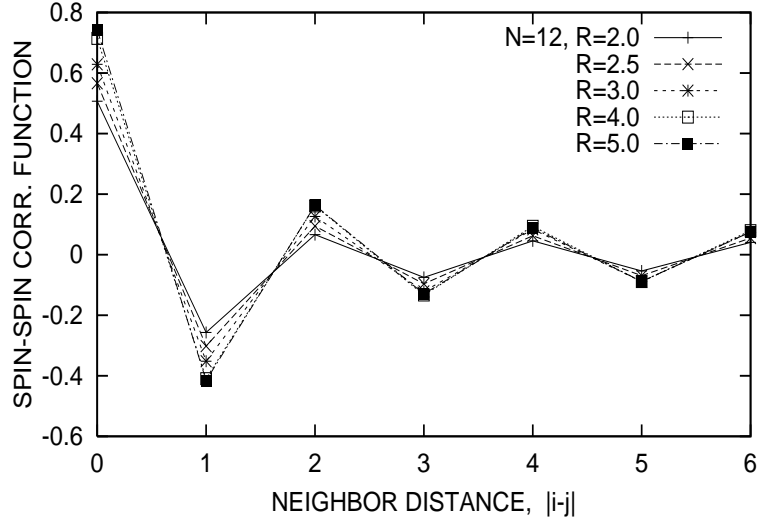}}
\caption{Spin-spin correlation function $\langle {\bf S}_i \cdot
{\bf S}_j \rangle$ vs. the distance $|i-j|$ between the atomic
sites, for different lattice constant $R$ and for $N = 12$ atoms.
Due to periodic boundary conditions only the distance up to
$|i-j|=6$ is relevant. The continuous line is guide to the eye. A
quasi-antiferromagnetic arrangement is clearly seen, particularly
for larger $R$.} \label{F5b}
\end{figure}

\begin{table}[!p]
\begin{center}
\begin{tabular}{r|lllll|ll}
\hline\hline $a/a_0$ & $D^*_{14}$ & $D^*_{12}$ & $D^*_{10}$ &
$D^*_{8}$ & $D^*_6$ & $D^*_{\infty}$ & {\small
$\sigma(D_{\infty}^*)/D_{\infty}^*$} \\ \hline 1.5 & 0.9225 &
0.9420 & 0.9563 & 0.9727 & 0.9822
 & 0.8008 & 0.019 \\
1.6 & 0.8879 & 0.9162 & 0.9378 & 0.9612 & 0.9754
 & 0.7175 & 0.029 \\
1.7 & 0.8419 & 0.8817 & 0.9130 & 0.9459 & 0.9667
 & 0.6148 & 0.043 \\
1.8 & 0.7826 & 0.8365 & 0.8805 & 0.9256 & 0.9552
 & 0.4967 & 0.064 \\
1.9 & 0.7095 & 0.7794 & 0.8389 & 0.8992 & 0.9406
 & 0.3728 & 0.092 \\
2.0 & 0.6245 & 0.7105 & 0.7875 & 0.8660 & 0.9222
 & 0.2567 & 0.129 \\
2.1 & 0.5315 & 0.6310 & 0.7265 & 0.8254 & 0.8996
 & 0.1606 & 0.172 \\
2.2 & 0.4338 & 0.5431 & 0.6549 & 0.7755 & 0.8714
 & 0.0899 & 0.228 \\
2.3 & 0.3403 & 0.4523 & 0.5766 & 0.7179 & 0.8379
 & 0.0455 & 0.287 \\
2.4 & 0.2554 & 0.3631 & 0.4937 & 0.6524 & 0.7982
 & 0.0207 & 0.352 \\
2.5 & 0.1839 & 0.2812 & 0.4109 & 0.5813 & 0.7526
 & 0.0087 & 0.420 \\
2.6 & 0.1269 & 0.2096 & 0.3315 & 0.5065 & 0.7009
 & 0.0033 & 0.489 \\
2.7 & 0.0840 & 0.1508 & 0.2595 & 0.4312 & 0.6441
 & 0.0012 & 0.566 \\
2.8 & 0.0536 & 0.1049 & 0.1972 & 0.3586 & 0.5836
 & 0.0004 & 0.641 \\
2.9 & 0.0315 & 0.0706 & 0.1456 & 0.2914 & 0.5208
 & 0.0001 & 0.920 \\ \hline
3.0 & 0.0196 & 0.0461 & 0.1047 & 0.2314 & 0.4575
 & 0.0000 & $\ \ \ -$ \\
\hline\hline
\end{tabular}
\end{center}
\caption{Normalized Drude weight $D_N^*$, the extrapolated value
$D_{\infty}^*$, and its relative error for 1D half--filled system
with long--range Coulomb interactions.} \label{T3}
\end{table}
In the previous Sections we illustrated the applications of the
EDABI method, in which {\em the interaction among particles is
dealt with first}. This is because, in most cases, the interaction
parameters (coupling constants) represent the largest energy scale
in the system. The first of the examples ({\em the Hubbard chain})
represents the situation, for which an analytic expression for the
ground state energy exists \cite{C12}, whereas the case of $H_N$
{\em rings} must be treated numerically all the way through
\cite{C18}. In applications of this method to the extended
three-dimensional systems one will have to resort to the
approximate treatments of the model Hamiltonian in the Fock space.
This last problem poses a real challenge for the future. In the
remaining part of this brief review we concentrate on the
collective properties of the nanochain.\\
\indent  We have concentrated first on the basic
quantum-mechanical features of the system such as the ground-state
energy or the renormalized single-particle wave function in the
milieu of other particles. In Fig. \ref{F5a} we plot the exact
ground state energy of a chain of $N=6 \div 10$ atoms and compare
it with that obtained in the Hartree-Fock approximation for the
Slater antiferromagnetic state. The starting Hamiltonian is of the
form (\ref{E2}). As one can see the Hartree-Fock energy represents
an upper estimate, as it should be. Additionally, the curve M
represents the "metallic" approximation, for which the correlation
function $<\delta n_i\delta n_j>$ has been taken for the 1D
electron gas on the lattice. On the contrary, INS represents the
energy of the Heisenberg-Mott state in the mean-field
approximation. The state of the system crosses over from the
Slater metallic-type state to the localized-spin-type of state.
This is seen explicitly when we calculate the evolution of the
spin-spin correlation function with the increasing interatomic
distance, as displayed in Fig. \ref{F5b}. Well defined
oscillations of $<{\bf S}_i\cdot{\bf S}_j
>$ are seen for even ($N=12$) number of atoms, which become more pronounced with the increasing $N$ (the frustration effects
appear for odd $N$). What is much more important, the
autocorrelation part $<{\bf S}_i\cdot{\bf S}_i> = <{\bf
S}_i^2>=(3/4)(1-2<n_{i\uparrow}n_{i\downarrow}>)$ evolves from the
value close to the free-electron value $<{\bf
S}_i^2>=(3/4)(1-2<n_{i\uparrow}>< n_{i\downarrow}>) = 3/8$ to the
atomic-limit value $<{\bf S}_i^2>=(1/2)(1/2+1)=3/4$. This
evolution provides a direct evidence of the crossover from
delocalized to
the localized regime.\\
\begin{figure}
\centerline{\includegraphics[width=10cm,height=6cm]{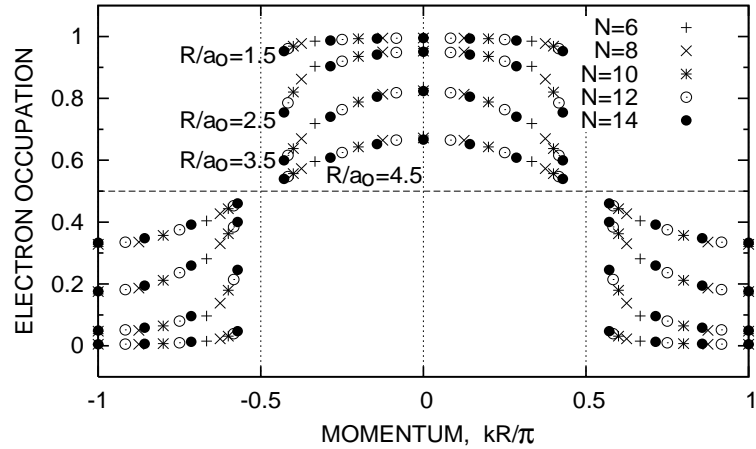}}
\caption{\protect\inx{Evolution} of the statistical momentum
distribution $n_{k\sigma}$ with the increasing interatomic
distance $R$. For the nanochain of $N = 6,10,$ and $14$ the
periodic boundary conditions provide the energy minimum, whereas
for $N = 8$ and $12$ antiperiodic boundary conditions are
appropriate. For details see main text.} \label{F6}
\end{figure}
\begin{figure}[!p]
\setlength{\unitlength}{0.01\textwidth}
\begin{picture}(100,110)
\put(-22,-18){\includegraphics[width=1.5\textwidth]%
{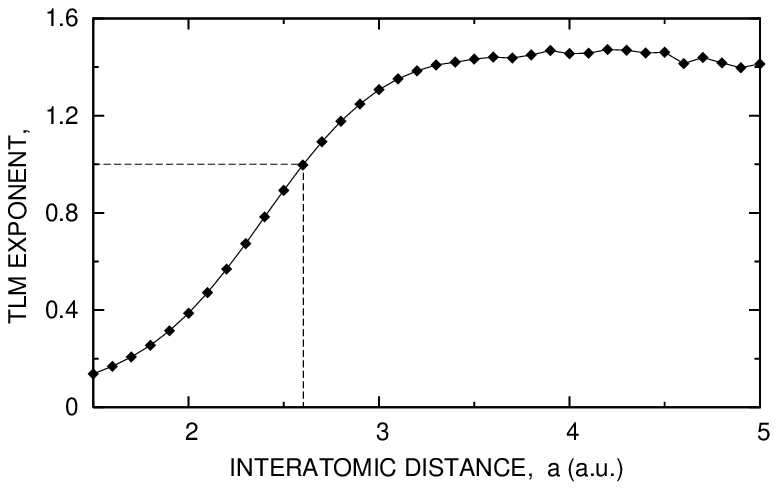}}
\put(-33,45){\includegraphics[width=1.25\textwidth]%
{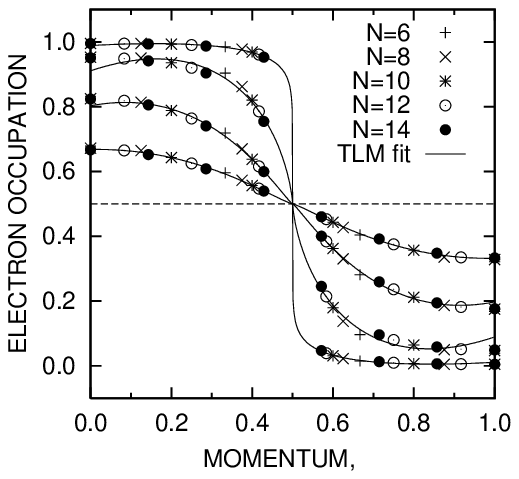}}
\put(18,45){\includegraphics[width=1.25\textwidth]%
{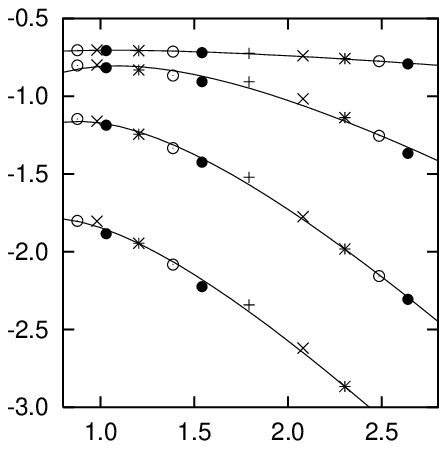}}
\put(7,-10){\includegraphics[width=1.25\textwidth]%
{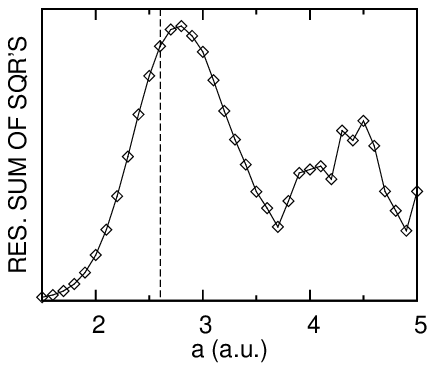}} \put(2.7,46.2){\rotatebox{90}{\large $\theta$}}
\put(71,63){$\ln (\pi/|k_F-k|a)$} \put(53,81){\rotatebox{90}{$\ln
|n_F-n_{{\bf k}\sigma}|$}} \put(20,106){\rotatebox{350}{$a=1.5$}}
\put(14,101){\rotatebox{345}{$a=2.5$}}
\put(14,97){\rotatebox{335}{$a=3.5$}}
\put(12,92){\rotatebox{345}{$a=4.5$}}
\put(86,102){\rotatebox{353}{$a=1.5$}}
\put(86,96){\rotatebox{338}{$a=2.5$}}
\put(83,86){\rotatebox{324}{$a=3.5$}}
\put(78,78){\rotatebox{318}{$a=4.5$}} \put(37,62.3){$ka/\pi$}
\put(11,71){(\textbf{a})} \put(63,71){(\textbf{b})}
\put(14,55){(\textbf{c})}
\end{picture}\caption{Luttinger--liquid scaling for a \emph{half--filled} 1D chain
of $N=6\div 14$ atoms with \emph{long--range} Coulomb
interactions: $(a)$ momentum distribution for electrons in the
linear and $(b)$ log--log scale, continuous lines represent the
fitted singular expansion in powers of $\ln(\pi/|k_F-k|a)$ (see
main text for details); $(c)$ Tomonaga--Luttinger model exponent
$\theta$ vs.\ lattice parameter $a$ (specified in $a_0$) and (in
the \emph{inset}) the corresponding residual sum of squares. The
solid lines in Figures $(a)$ and $(b)$ represent the TLM fitting
of Eq.\ (\ref{E9}).} \label{F10}
\end{figure}
\indent Other properties such as the electrical conductivity
\cite{C16} and the statistical distribution in momentum space
($n_{k\sigma}$) have also been addressed \cite{C9}. Here the
question emerges whether the {\em quantum nano-liquid} of
electrons in a nanochain resembles at all the Landau-Fermi liquid
or if it is rather represented by the Tomonaga-Luttinger scaling
laws \cite{C17}. The answer is not yet settled, as within our
method we can deal only with up to N=16 hydrogen atoms assembled
into a linear chain. However, one can make some definite
statements for the particular cases. Namely, for the half-filled
case (one electron per atom) the modified Fermi distribution is a
good representation of the $n_{k\sigma}$ for smaller $R$ values;
with the increasing atom spacing it is smeared out above critical
spacing $R=R_c\approx3.4 a_0$ \cite{C10}, which characterizes a
{\em crossover} from the case with extended states to the state of
localized electrons on atoms, as detailed below. In Fig. \ref{F6}
we exhibit this evolution on the example of the distribution
function (i.e. electron occupation in the momentum space); a clear
universality is observed for $N=6 \div 14$ atoms. The adjustable
Gaussian (STO-3G) single-particle basis has been used in the
analysis. The existence of the Fermi ridge quasi-discontinuity for
a small $R$ is very suggestive in this case and is positioned near
the Fermi wave-vector $k_F^\infty = \pm \pi / (2 R)$,
corresponding to that in Landau-Fermi liquid, with $N \rightarrow
\infty$. For $N= 6, 10, 14$ periodic boundary conditions (PBC)
provide the minimal ground state energy, whereas for $N=8$ and
$12$  the anti-periodic boundary conditions (ABC) lead to the
lower energy. However, one clearly sees the absence of the points
at the Fermi momentum $k_F^\infty$. This is because, for example,
the Fermi points for $N=14$ are located at $k_FR/\pi = \pm 3/7$,
whereas they are located at $k_FR/\pi = \pm 5/12$ for $N=12$,
close to the values $\pm 1/2$ in both cases. One would have to
apply a renormalization group approach \cite{C19} for the states
close to $k_F$ (i.e. perform the analysis for larger number $N
\sim 10^2$ atoms) to determine the precise evolution of the
distribution function, this time with the system size $N$.
Nevertheless, the results for $N \leq 14$ atoms represent those
for a true nanoscopic system.\\
\indent Before addressing the question of localization directly,
we would like to address the question whether the computed
distribution displayed in Fig. \ref{F6} can be fitted into the
Tomonaga-Luttinger mode, with the logarithmic scaling corrections
included \cite{C17}. The statistical distribution near the Fermi
point can be represented by
\begin{equation}
\ln |n_{k \sigma}-n_F| = -\theta \ln z + b \ln\ln z + O(1/\ln z),
\label{E9}
\end{equation}
with $z \equiv \pi / |k - k_F|$. Here $\theta$ is a non-universal
(interaction-dependent) exponent (it yields the nonexistence of
fermionic quasiparticles,  since its residue vanishes as $Z_k \sim
|k - k_F|^\theta$ with $k \rightarrow k_F$. The corresponding
electron-momentum distribution  is depicted in Fig. \ref{F10}a in
the linear, and in Fig. \ref{F10}b in the log-log scale. The $R$
dependence of the exponent $\theta$ is shown in Fig. \ref{F10}c
and crosses the value $\theta = 1$ for $R = R_c \approx 2.6 a_0$
corresponding to the localization threshold \cite{C17}. This
threshold  is about 30\% smaller than the corresponding value
($R_c \approx 3.4 a_0$) when the almost-localized Fermi-liquid
view was taken \cite{C10}. One should also note that the Luttinger
scaling does not reproduce well the occupancies farther away
either way from the Fermi point. Thus the results concerning
$n_{k\sigma}$ do not provide a definite answer as to the exact
nature of the {\em nanoliquid} composed of $N \sim 10$ electrons,
although it is absolutely amazing that they have such a nice and
simple scaling properties.\\
\indent To address the question of the electron localization
directly, we have calculated the optical conductivity
$\sigma(\omega)$, which can be written in the form $\sigma(\omega)
= D\delta(\omega) + \sigma_{reg}(\omega)$, where the regular part
is
\begin{equation}
\sigma_{reg}(\omega) \frac{\pi}{N} \sum_{n \neq 0} \frac{|< \Psi_n
| j_p | \Psi_0 >|^2}{E_n - E_0} \delta(\omega - ( E_n - E_0 )),
\end{equation} whereas the {\em Drude weight (the charge
stiffness)} $D$ is given by
\begin{equation}
D = \frac{\pi}{N} < \Psi_0 | T | \Psi_0 >  - \frac{2\pi}{N}
\sum_{n \neq 0} \frac{|< \Psi_n | j_p | \Psi_0 >|^2}{E_n - E_0},
\end{equation} with $T$ being the hopping term as in (\ref{E1}) and
$j_p$ the current operator defined as $j_p = i t \sum_{i j \sigma}
(a^\dagger_{j \sigma} a_{i \sigma} - h.c.)$. Here $| \Psi_n >$ is
the system eigenstate corresponding to the eigenvalue $E_n$. For a
finite system of N atoms D is always nonzero due to nonzero
tunnelling rate  through a potential barrier of finite width.
Because of that, the finite-size scaling with $1/N \rightarrow 0$
must be performed on $D$. We use the following parabolic
extrapolation
\begin{equation}
\ln D^\star_N = a + b (1/N) + c (1/N)^2, ]
\end{equation}
where $D^\star_N = -(N/\pi) D/ <\Psi_0| T |\Psi_0>$ denotes the
normalized Drude weight for the system of N sites. We observe that
$0 \leq D^\star \leq 1$ and hence can be regarded as an order
parameter for the transition to the localized (atomic) states. In
Table \ref{T3} we plot the weights $D^\star_N$, the extrapolated
values $D^\star_\infty$, and its relative error for 1D half-filled
system with long-range Coulomb interactions included. What is very
important, the value of $D^\star_N$ drops by two orders of
magnitude when R changes by a factor of two (between $1.5 a_0$ and
$3 a_0$). Note also that $D^\star_\infty$ is within its error for
$R \simeq R_c \approx 2.3 a_0$, close to the value obtained from
the Tomonaga-Luttinger scaling for $n_{k \sigma}$, as one would
expect. The result for $D^\star_N$ is probably telling us how far
we can go {\em quantitatively} when discussing the localization in
nanowires. These results will be detailed in a separate
publication.
 \subsection{Quarter-filled case}
\begin{figure}
\setlength{\unitlength}{0.01\textwidth}
\begin{picture}(100,54)
\put(-22,-24){\includegraphics[width=1.5\textwidth]{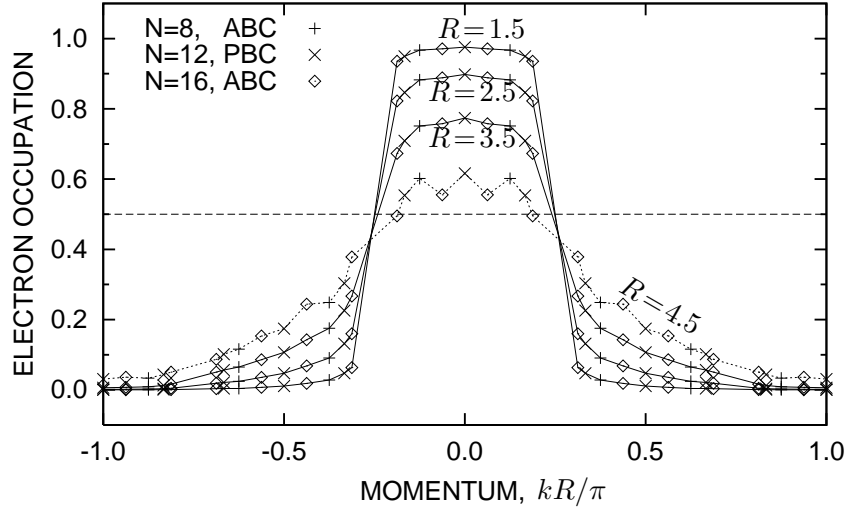}}
\put(50,48){$R\!=\!1.5$} \put(49,41){$R\!=\!2.5$}
\put(49,36){$R\!=\!3.5$} \put(70,20){\rotatebox{330}{$R\!=\!4.5$}}
\put(61,-3){$kR/\pi$}
\end{picture}
\vspace{0.5em} \caption{Momentum distribution $n_{k\sigma}$ for
electrons on a chain of $N=8\div 16$ atoms in the
\emph{quarter--filled} band case ($N_e=N/2$). Lines are drawn as a
guide to the eye only. Values of the lattice parameter $R$ are
specified in units of $a_0$. PBC and ABC denote periodic and
antiperiodic boundary conditions, respectively. The dashed line
marks occupation $n_{k\sigma} = 1/2$.} \label{F7}
\end{figure}
\begin{figure}
\setlength{\unitlength}{0.01\textwidth}
\begin{picture}(100,45)
\put(-8,-20){\includegraphics[width=1.25\textwidth]{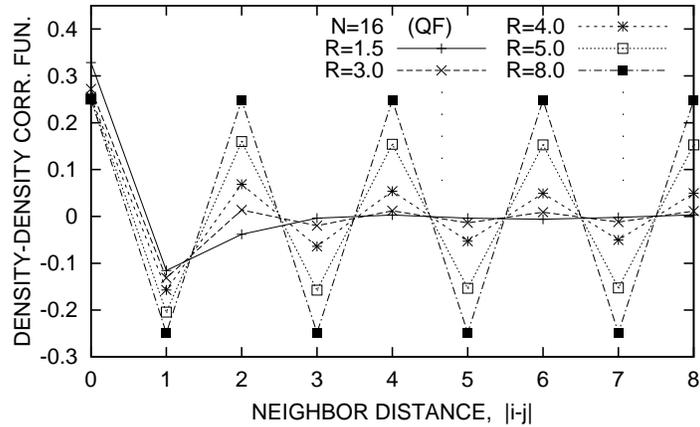}}
\end{picture}
\vspace{0.5em} \caption{Charge--density distribution for the
\emph{quarter--filled} ($N_e=N/2$) nanochain as viewed by
density--density fluctuation correlation function $\left<\Delta
n_i\Delta n_j\right>$ vs.\ relative distance $\left|i-j\right|$.}
\label{F8}
\end{figure}
For the quarter-filled case (i.e. when every second atom
contributes a valence electron) the distribution ${n}_{{\bf k}
\sigma}$ is more smeared out, as shown in Fig. \ref{F7}. The
ground state for $R \geq 3 a_0$ is then well represented by the
charge-density-wave state, since then the density-density
correlation function
$\langle(n_i-\overline{n})(n_j-\overline{n})\rangle$ exhibits the
oscillatory behavior, as shown in Fig. \ref{F8}. The onset of the
charge-density wave state in that case is invariably due to the
long-range Coulomb interaction $\sim K_{ij}$, which is reduced in
that state. The charge-density wave order parameter defined as
\begin{equation}
\Theta_{CDW} \equiv \frac{1}{N} \sum_{m} (-1)^m
<(n_i-\overline{n})(n_j-\overline{n})>
\end{equation}
reaches its maximal value $1/4$ for $R \geq 8 a_0$.
 Let us stress again, the form
of the statistical distribution (and its feasibility) for $N \sim
10$ atoms is a very interesting question by itself, since this is
{\em the regime of nanoscience}. Our results show that even in
that regime one should be able to see the signatures of the phase
transition to the spin- or charge- density wave states.
\section{Conclusions}
The EDABI method ({\bf E}xact {\bf D}iagonalization combined
with{\em {\bf AB}  {\bf I}nitio} orbital readjustment)provides the
{\em exact} ground state energy of the model systems considered
({\em Hubbard chain, nanoscopic chains and rings}) as a function
of interatomic distance. It also provides reliably other
ground-state dynamical characteristics for nanoscopic systems:
spin and charge correlation functions, the spectral density and
the density of states, as well as the system conductivity. Not all
the characteristics have been presented in this overview
\cite{C5,C10,C16,C17}. Furthermore, the method can also be
extended to nonzero temperatures. Finally, it should be underlined
again that our method of approach is particularly suited for
strongly correlated systems, where the interaction and the
single-particle parts should be treated on the same footing. The
exact numerical results in a model situation can also serve as a
test for approximate analytic treatments.\\
\indent The analysis with inclusion of long-range Coulomb
interactions of the distribution function $n_{k \sigma}$ in either
Fermi-like or Tomonaga-Luttinger categories suggests the existence
of the {\em crossover} transition to the localized state with the
increasing interatomic distance $R$. This state is either the
spin- or the charge- ordered for the number of electrons $N_e = N$
and $N/2$, respectively. In the small $R$ limit the state can be
rendered as quasi-metallic in the sense that the quasimomentum
$\hbar k$ can be regarded as a good quantum number, even though
the level structure is discrete. The difference between the
short-chain and infinite-chain situation is due to the
circumstance that in order to form extended states in the present
case the electrons tunnel through a barrier of finite width (the
length $L = NR$). This is one of the reasons for
quasi-metallicity. The other is the presence of the long-range
Coulomb interaction\cite{C20}.
\section{Acknowledgment}
The work was supported by the State Committee for Scientific
Research KBN, Grant No. 2P03B 050 23. The two authors (A.R \&
J.S.) acknowledge respectively the junior and the senior
fellowships of the Foundation for Science (FNP). We are also
grateful to the Institute of Physics of the Jagiellonian
University  for the support for computing facilities used in part
of the numerical analysis.






%





\begin{chapthebibliography}{xxx}
\bibitem{C11} If we want to include it, then it is irrelevant for the case of one electron per atom, since then pure spin-density-wave correlations set in.
However, if the system is non-neutral ($N_e \neq N$), then the
charge-density-wave state may become stable (c.f. Fig. \ref{F8} in
Sec. 5).
\end{chapthebibliography}
\end{document}